\documentclass[transaction]{IEEEtran}
\usepackage{graphicx}
\usepackage{graphicx}
\usepackage{epstopdf}
\usepackage[latin1]{inputenc}
\usepackage{amsmath,amssymb,amscd,latexsym,dsfont}
\usepackage[english]{babel}
\usepackage{tabularx,cite}
\usepackage{graphicx}
\usepackage{algpseudocode}
\usepackage{algorithm}
\usepackage{algorithmicx}
\usepackage{float}
\usepackage{multicol}
\usepackage{psfrag}
\usepackage{comment,psfrag,subfigure,enumerate}
\usepackage{color}
\usepackage{amsthm}
\ifCLASSINFOpdf
\else
\fi
\begin{document}
%
\title{HARQ Feedback in Spectrum Sharing Networks}
\author{\IEEEauthorblockN{Behrooz Makki, Alexandre Graell i Amat, \emph{Senior Member, IEEE,} and Thomas Eriksson}\\
\thanks{Alexandre Graell i Amat was supported by the Swedish Agency for Innovation Systems (VINNOVA) under the P36604-1 MAGIC project.}
\thanks{Behrooz Makki, Alexandre Graell i Amat and Thomas Eriksson are with the Department of Signals and Systems,
Chalmers University of Technology, Gothenburg, Sweden, Email: \{behrooz.makki, alexandre.graell, thomase\}@chalmers.se}
}

%
\maketitle
\vspace{-15mm}
\begin{abstract}
This letter studies the throughput and the outage probability of spectrum sharing networks utilizing hybrid automatic repeat request (HARQ) feedback. We focus on the repetition time diversity and the incremental redundancy HARQ protocols where the results are obtained for both continuous and bursting communication models. The channel data transmission efficiency is investigated in the presence of both secondary user peak transmission power and primary user received interference power constraints. Finally, we evaluate the effect of secondary-primary channel state information imperfection on the performance of the secondary channel. Simulation results show that, while the throughput is not necessarily increased by HARQ, substantial outage probability reduction is achieved in all conditions.
\end{abstract}
%
\IEEEpeerreviewmaketitle
%
%
%
%
\vspace{-6mm}
%
%
\section{Introduction}
\vspace{-0mm}
To tackle today's spectrum shortage problem, several solutions have been proposed among which spectrum sharing is one of the most promising ones \cite{5723046,cogasli,revisioncogref3,ICC1,compressedcogHARQ2,revisioncogref2}. In a spectrum sharing network, unlicensed secondary users (SUs) are permitted to work within the spectrum resources of licensed primary users (PUs) as long as the PUs quality-of-service requirements are satisfied.

Hybrid automatic repeat request (HARQ) is an efficient approach for increasing the data transmission efficiency of different communication setups. Utilizing HARQ in spectrum sharing networks has recently attracted considerable attention, e.g., \cite{revisioncogref2,revisioncogref3,compressedcogHARQ2,ICC1}. In \cite{revisioncogref3,ICC1}, the SU works as a relay helping the PU, which uses incremental redundancy (INR) HARQ. Also, in \cite{revisioncogref2,compressedcogHARQ2} the INR HARQ is exploited by the PU for increasing its protection against the SU interferences.

In this paper, as opposed to \cite{revisioncogref2,revisioncogref3,compressedcogHARQ2,ICC1} where the use of INR HARQ is limited to the PU, we consider both the repetition time diversity (RTD) and the INR HARQ protocols in the secondary channel to study the data transmission efficiency of spectrum sharing networks. The goal is to analyze the SU-SU channel throughput and the SU outage probability under PU received interference power and SU peak transmission power constraints. Considering block fading channels, the results are obtained for both continuous and bursting communication models in the case where the SU transmitter is provided with imperfect SU-PU channel state information (CSI). The results indicate that implementation of HARQ does not necessarily increase the throughput. However, substantial outage probability reduction is achieved by HARQ in all conditions.

\vspace{-5mm}
\section{System model}
Consider a block fading spectrum sharing network where a PU shares the same narrow-band frequency with an unlicensed SU. Let $h_\text{pp}$, $h_\text{ps}$, $h_\text{sp}$ and $h_\text{ss}$ be the fading random variables in the PU-PU, PU-SU, SU-PU and SU-SU links, respectively. Correspondingly, we define the \emph{channel gains} ${g_{\text{pp}}} \buildrel\textstyle.\over= |{h_{\text{pp}}}{|^2}$, ${g_{\text{ps}}} \buildrel\textstyle.\over= |{h_{\text{ps}}}{|^2}$, ${g_{\text{sp}}} \buildrel\textstyle.\over= |{h_{\text{sp}}}{|^2}$ and ${g_{\text{ss}}} \buildrel\textstyle.\over= |{h_{\text{ss}}}{|^2}$. The channel gains remain constant for a duration of $L_\text{c}$ channel uses, generally determined by the channel coherence time, and then change independently according to the fading probability density functions (pdfs) ${f_{g_\text{pp}}}(g)$, ${f_{g_\text{ps}}}(g)$, ${f_{g_\text{sp}}}(g)$, ${f_{g_\text{ss}}}(g)$, respectively. The simulations are focused on Rayleigh fading channels, ${f_{{g}}}(x) = \frac{1}{{{\mu}}}{e^{ - \frac{x}{{{\mu}}}}}$, where $\mu$ represents the fading parameter determined based on the path loss and shadowing between the terminals. Finally, the AWGN at the PU and the SU receivers is assumed to have independent and identically distributed complex Gaussian distribution $\mathcal{CN}(0,N_0)$. Also, in harmony with, e.g., \cite{5723046,cogasli,revisioncogref3,ICC1,compressedcogHARQ2,revisioncogref2}, the PU transmission signal is supposed to have Gaussian pdf with power $P_\text{p}$, which leads to AWGN interference at the SU receiver.

We assume perfect CSI about the SU-SU and PU-SU channel gains at the SU receiver, which is an acceptable assumption in block fading channels \cite{revisioncogref2,ICC1,throughputdef,5336856}. Also, the SU transmitter is provided with some imperfect CSI of the SU-PU channel modeled by
\vspace{-2mm}
\begin{align}
{{\tilde h}_\text{{sp}}} = \beta {h_\text{{sp}}} + \sqrt {1 - {\beta ^2}} \varepsilon,\, \varepsilon \sim \mathcal{CN}(0,\mu_\text{sp}),\,0\le \beta\le 1
\vspace{-1mm}
\end{align}
where ${{\tilde h}_\text{{sp}}}$ is the SU-PU channel estimate provided at the SU transmitter, $\beta$ is a known correlation factor modeling the estimation quality, $\mu_\text{sp}$ is the SU-PU fading parameter and $\varepsilon$ is a complex Gaussian variable independent of ${{ h}_\text{{sp}}}$. This is a well accepted model for partial CSI \cite{cogasli,compressedneweurasipkhodemun,cog2}.

A maximum of $M$ retransmission rounds are considered, i.e., the data is (re)transmitted by the SU a maximum of $M+1$ times until it is successfully decoded by the SU receiver or the maximum number of rounds is reached. Both the RTD and the INR protocols are implemented for the HARQ. Finally, the feedback bits are assumed to be delivered at the SU transmitter error- and delay-free.
\vspace{-5mm}
\section{System throughput}
\vspace{-1mm}
We call the transmission of a codeword along with all its possible retransmission rounds a \emph{packet}. The long-term throughput (in nats-per-channel-use (npcu)) is defined as \cite{throughputdef}
\vspace{-2mm}
\begin{align}
\eta  \buildrel\textstyle.\over= \mathop  \frac{{\bar D }}{{\bar \l  }}
\vspace{-7mm}
\end{align}
where $\bar D$ and $\bar l$ denote the expected value of the successfully-decoded information nats\footnote{All results are presented in natural logarithm basis.} and the total number of channel uses within a fading block, respectively.

Both continuous and bursting communication schemes \cite{5336856} are considered. Under the continuous communication model, it is assumed that there is an infinite amount of information available at the SU transmitter and it is always active. Thus, multiple packets, each packet containing multiple HARQ rounds, are transmitted within one fading block of length $L_\text{c}$. If the channel is \emph{good}, many packets are sent within a fading block, while only few can be transmitted within the same period for \emph{bad} channels. In this case, the long-term throughput is calculated as follows. Let $R(g_\text{ss},g_\text{ps})$ be the SU instantaneous data rate of the HARQ approach for given gain realizations $g_\text{ss}$ and $g_\text{ps}$. Then, the total number of information nats decoded in each state is obtained by $D(g_\text{ss},g_\text{ps})=L_\text{c} R(g_\text{ss},g_\text{ps})$. Consequently, the long-term throughput is given by
\vspace{-1mm}
\begin{align}
\eta  = \frac{{E\{ {L_\text{c}}R(g_\text{ss},g_\text{ps})\} }}{{{L_\text{c}}}} = E\{ R(g_\text{ss},g_\text{ps})\}  = \bar R,
\vspace{-5mm}
\end{align}
i.e., the channel average rate.

Under the bursting communication model, on the other hand, it is assumed that there is a long idle period between the transmission of two packets. Therefore, while the HARQ retransmission rounds of each packet experience the same gains realizations, the channels change independently from one packet to another. In this case, the denominator of (2) is not constant and, as discussed in the sequel, should be calculated separately. More specifically, as opposed to the continuous communication model, where all the $L_\text{c}$ channel uses of a fading block are utilized, in the bursting communication model only one packet is sent within each block that, depending on the channels conditions, can be
decoded by the SU receiver in different (re)transmission rounds.


Let $A_m$ be the event that the data is successfully decoded at the $m$-th, $m=1,\ldots,M+1$, (re)transmission round of the HARQ protocol and not before. In this way, the system throughput under the continuous communication assumption is obtained by
\vspace{-2mm}
\begin{align}
\eta  = \sum_{m = 1}^{M + 1} {{R_m}\Pr \{ {A_m}\} }
\vspace{-2mm}
\end{align}
where $R_m$ represents the equivalent data rate after $m$ (re)transmission rounds. Also, the data is lost and an outage happens if the data can not be decoded after $M+1$ (re)transmission rounds. Therefore, the outage probability is found as
\vspace{-2mm}
\begin{align}
\Pr {\{ \text{outage}\} } = 1 - \sum_{m = 1}^{M + 1} {\Pr \{ {A_m}\} }.
\vspace{-6mm}
\end{align}

To find the system throughput under the bursting communication model, assume that $D$ information nats are transmitted in each packet transmission. Provided that the data is decoded at any (re)transmission round, all the $D$ nats are received by the SU receiver. Therefore, the expected number of received information nats in each packet is
\vspace{-2mm}
\begin{align}
\bar D  = D\left(1 - \Pr \{ \text{outage}\} \right).
\vspace{-3mm}
\end{align}
If the data (re)transmission successfully stops at the $m$-th (re)transmission round the total number of channel uses is $\sum_{n=1}^{m}{l_n}$, where $l_n$ is the length of the codeword sent in the $n$-th (re)transmission round. Also, there will be $\sum_{n=1}^{M+1}{l_n}$ channel uses if an outage happens, as all possible (re)transmission rounds are used. Hence, the expected number of channel uses within a packet is found as
\vspace{-2mm}
\begin{align}
\bar l  = \sum_{m = 1}^{M + 1} {\left(\sum_{n = 1}^m {{l_n}} \right)\Pr \{ {A_m}\}  + \left(\sum_{n = 1}^{M + 1} {{l_n}} \right)\Pr \{ \text{outage}\} }
\vspace{-2mm}
\end{align}
and the throughput in the bursting communication model is
\vspace{-2mm}
\begin{align}
\eta  = \frac{{D(1 - \Pr \{ \text{outage}\} )}}{{\sum_{m = 1}^{M + 1} {(\sum_{n = 1}^m {{l_n}} )\Pr \{ {A_m}\}  + (\sum_{n = 1}^{M + 1} {{l_n}} )\Pr \{ \text{outage}\} } }}.
\vspace{-2mm}
\end{align}
In the following, (4)-(8) are studied in more detail for both the RTD and the INR HARQ protocols.
\vspace{-4mm}
\subsection{RTD protocol}
Using RTD, $D$ information nats are encoded in each codeword of length $L,\, {L} \ll {L_\text{c}}$, i.e., the initial transmission rate is $R=\frac{D}{L}$. The same codeword is retransmitted in the successive retransmission rounds. Hence, the equivalent transmission rate at the end of the $m$-th (re)transmission round is $R_m=\frac{D}{mL}=\frac{R}{m}$. Also, the receiver performs maximum ratio combining of the received signals. Thus, the SU received signal-to-interference-and-noise ratio (SINR) in the $m$-th (re)transmission round is
\vspace{-1mm}
\begin{align}
{\gamma _m} = m\Omega ,\,\,\Omega  \buildrel\textstyle.\over= \frac{{{P_\text{s}}{g_{\text{ss}}}}}{{{P_\text{p}}{g_{\text{ps}}} + {N_0}}}
\end{align}
where $P_\text{s}$ denotes the SU transmission power. Therefore, the probability term $\Pr\{A_m\}$ is obtained by\footnote{In (10), we have used the fact that with an equivalent SINR $x$ the maximum decodable transmission rate is $\frac{1}{m} \log(1+x)$ if a codeword is repeated $m$ times.}
\begin{align}
\begin{array}{l}
 \Pr {\{ {A_m}\} } = \Pr \left\{ \log (1 + (m - 1)\Omega ) \le R < \log (1 + m\Omega )\right\}  \\
  \,\,\,\,\,\,\,\,\,\,\,\,\,\,\,\,\,\,\,\,\,\,\,\,= {F_\Omega }(\frac{{{e^R} - 1}}{{m - 1}}) - {F_\Omega }(\frac{{{e^R} - 1}}{m}) \\
 \end{array}
\end{align}
and the outage probability is found as
\vspace{-2mm}
\begin{align}
\Pr \{ \text{outage}\}  =\Pr \left\{ \log (1 + (M + 1)\Omega ) < R\right\} = {F_\Omega }(\frac{{{e^R} - 1}}{{M + 1}}).\nonumber
\vspace{-2mm}
\end{align}
Here, ${F_\Omega }$ is the cumulative distribution function (cdf) of the auxiliary variable $\Omega$ obtained based on the PU and the SU quality-of-service requirements (see Subsection III.C). Considering $R_m=\frac{R}{m}$ and (10) the throughput in the continuous communication model (4) is obtained easily. On the other hand, as we have $R=\frac{D}{L}$ and $l_m=L\, \forall m$, the throughput under the bursting communication assumption (8) is rephrased as
\vspace{-1mm}
\begin{align}
\eta  = \frac{{R(1 - \Pr \{ \text{outage}\} )}}{{\sum_{m = 1}^{M + 1} {m\Pr \{ {A_m}\}  + (M + 1)\Pr \{ \text{outage}\} } }}.
\vspace{-2mm}
\end{align}
\vspace{-6mm}
\subsection{INR protocol}
In the INR scheme, new variable-length codewords are sent in the successive (re)transmission rounds of a packet. Then, in each (re)transmission round the message is decoded by the SU receiver using all previously received signals of the packet. Hence, denoting the equivalent transmission rate at the end of the $m$-th (re)transmission round by $R_m=\frac{D}{\sum_{n=1}^{m}{l_n}}$, the probability term $\Pr\{A_m\}$, the throughput in the continuous communication model, and the outage probability are
\vspace{-1mm}
\begin{align}
\begin{array}{l}
 \Pr {\{ {A_m}\} } = \Pr \left\{ {R_m} \le \log (1 + \Omega ) < {R_{m - 1}}\right\}  \\
  \,\,\,\,\,\,\,\,\,\,\,\,\,\,\,\,\,\,\,\,\,\,\,\,= {F_\Omega }({e^{{R_{m - 1}}}} - 1) - {F_\Omega }({e^{{R_m}}} - 1), \\
 \end{array}
 \vspace{-2mm}
\end{align}
\vspace{-4mm}
\begin{align}
\eta  = \sum_{m = 1}^{M + 1} {{R_m}\left({F_\Omega }\left({e^{{R_{m - 1}}}} - 1\right) - {F_\Omega }\left({e^{{R_m}}} - 1\right)\right)},
\vspace{-2mm}
\end{align}
\vspace{-4mm}
\begin{align}
\Pr \{ \text{outage}\}  =\Pr \{ \log (1 + \Omega ) < {R_{M + 1}}\}= {F_\Omega }({e^{{R_{M + 1}}}} - 1),\nonumber
\vspace{-2mm}
\end{align}
respectively, and the throughput in the bursting communication model is
\vspace{-2mm}
\begin{align}
\eta  = \frac{{1 - \Pr \{ \text{outage}\} }}{{\sum_{m = 1}^{M + 1} {\frac{1}{{{R_m}}}\Pr \{ {A_m}\}  + \frac{1}{{{R_{M + 1}}}}\Pr \{ \text{outage}\} } }}.
\vspace{-2mm}
\end{align}
Here, (12) is based on the fact that the data is decoded at the end of the $m$-th (re)transmission round if 1) it has not been decoded before, i.e., $\log (1 + \Omega ) < {R_{m - 1}}<\ldots<R_1$, and 2) the equivalent transmission rate at the end of the $m$-th time slot is supported by the channel gains realizations, that is, ${R_m} \le \log (1 + \Omega )$.
\vspace{-3mm}
\subsection{Transmission power constraints}
We consider two simultaneous transmission power constraints; 1) the SU peak transmission power should be less than a threshold $P_\text{max}$, i.e., $P_\text{s}\le P_\text{max}$, and 2) the PU received interference power should not exceed a given value $I_\text{p}$. Therefore, the SU transmission power is selected as ${P_\text{s}} = \min (P_\text{max},\frac{{{I_\text{p}}}}{{{{\tilde g}_{\text{sp}}}}})$ where ${{\tilde g}_{\text{sp}}} = |{{\tilde h}_{\text{sp}}}{|^2}$ is the SU-PU channel estimate available at the SU transmitter. In this way, as illustrated in Appendix A, the cdf of the auxiliary variable $\Omega$, defined in (9), is found as
\vspace{-2mm}
\begin{align}
\begin{array}{l}
 {F_\Omega }(x) = 1 - \frac{{(1 - {e^{ - \frac{{{I_\text{p}}}}{{{\mu _{\text{sp}}}P_\text{max}}}}}){e^{ - \frac{{{N_0}}}{{{\mu _{\text{ss}}}P_\text{max}}}x}}}}{{1 + \frac{{{P_\text{p}}{\mu _{\text{ps}}}}}{{{\mu _{\text{ss}}}P_\text{max}}}x}} \\
  - \frac{{{\mu _{\text{ss}}}{I_\text{p}}}}{{{\mu _{\text{sp}}}{\mu _{\text{ps}}}{P_\text{p}}x}}{e^{(\frac{{{N_0}}}{{{P_\text{p}}{\mu _{\text{ps}}}}} + \frac{{{\mu _{\text{ss}}{I_\text{p}}}}}{{{\mu _{\text{sp}}}{\mu _{\text{ps}}}{P_\text{p}}x}})}} \Gamma \left(0,(\frac{1}{{{P_\text{p}}{\mu _{\text{ps}}}}} + \frac{x}{{{\mu _{\text{ss}}}P_\text{max}}})({N_0} + \frac{{{\mu _{\text{ss}}}{I_\text{p}}}}{{{\mu _{\text{sp}}}x}})\right) \\
 \end{array}
\end{align}
where $\Gamma(x,y)$ is the incomplete Gamma function.

As ${{\tilde g}_{\text{sp}}} \ne {g_{\text{sp}}}$, the PU received interference power $\phi_\text{p}= P_\text{s} g_\text{sp}$ may exceed the threshold $I_\text{p}$. However, using the PU received interference cdf
\vspace{-2mm}
\begin{align}
\begin{array}{l}
 {F_{\phi_\text{p}} }(x|I_\text{p}) = 1 - {e^{ - \frac{x}{{{\mu _{\text{sp}}}P_\text{max}}}}} + \frac{t}{r}Q(\sqrt {\frac{{(u - r)x}}{{2P_\text{max}}}} ,\sqrt {\frac{{(u + r)x}}{{2P_\text{max}}}} ) \\
  + {e^{ - \frac{x}{{{\mu _{\text{sp}}}P_\text{max}}}}}Q(\beta \sqrt {\frac{{2x}}{{P_\text{max}w}}} ,\sqrt {\frac{{2{I_\text{p}}}}{{P_\text{max}w}}} ) - \frac{1}{2}(1 + \frac{t}{r}){e^{\frac{{ - ux}}{{2P_\text{max}}}}}{I_0}(\frac{{2\beta \sqrt {x{I_\text{p}}} }}{{P_\text{max}w}}) \\
 w = {\mu _{\text{sp}}}(1 - {\beta ^2}),\,u = \frac{2}{{{\mu _{\text{sp}}}}}(1 + \frac{{{\beta ^2}{\mu _{\text{sp}}}}}{w} + \frac{{{I_\text{p}}{\mu _{\text{sp}}}}}{{xw}}) \\
 t = u - \frac{{4{I_\text{p}}}}{{wx}},\,r = \sqrt {{u^2} - \frac{{16{\beta ^2}{I_\text{p}}}}{{x{w^2}}}}  \\
 \end{array}\nonumber
\end{align}
(see Appendix A) one can find a new threshold $\hat I_\text{p} \le I_\text{p}$ such that the PU received interference threshold is satisfied with some probability $\pi$, i.e., $\Pr\{\phi_\text{p}<I_\text{p}\}\ge \pi$. The new threshold is found as the solution of $F_{\phi_\text{p}}(I_\text{p}|\hat I_\text{p})=\pi$. Also, since $Q(x,0)=1$ and $I_0(0)=1$, the interference cdf ${F_{\phi_\text{p}} }(x|I_\text{p})$ is rephrased as
\vspace{-2mm}
\begin{align}
{F_{{\phi _{\text{p}}}}}(x|{I_{\text{p}}}) = \frac{1}{2}\left(1 + \frac{t}{r}\right)
\end{align}
as ${P_{\max }} \to \infty $, i.e., under relaxed SU peak power constraint. Finally, the PU SINR constraint is not considered here, although it can be mapped to the interference constraint in some cases. Also, the PU SINR constraint is normally studied under (im)perfect PU-PU CSI assumption, which is not considered in our model.
\vspace{-4mm}
\section{Simulation results and discussions}
\vspace{-1mm}
For both protocols, the initial transmission rate is set to $R_1=0.5$. Also, we consider equal-length coding for the INR which, as $R_m=\frac{D}{\sum_{n=1}^{n=m}{l_n}}$, leads to $R_m=\frac{R_1}{m}$, i.e., the same rates as in the RTD scheme. Setting $P_{\text{max}}=2$ and $P_\text{p}=0.5$, Figs. 1a and 1b show the throughput and the outage probability of the considered protocols in the bursting and continuous communication models under perfect SU-PU CSI assumption ($\beta=1$). Here, the system performance with no HARQ feedback, i.e., $M=0$, is considered in the figures as a comparison yardstick. Note that the outage probability is the same in these communication models. Also, with proper scaling, the results can be mapped to the case with imperfect SU-PU CSI.

Assuming imperfect SU-PU CSI ($\beta=0.8$), Figs. 2a and 2b study the system throughput as a function of the PU received interference probability constraint $\pi$ and the PU transmission power $P_\text{p}$, respectively. Here, the results are obtained for a maximum of $M=1$ retransmission round and under different SU peak transmission power constraints. Finally, in all simulations the fading parameters are set to $\mu_\text{ss}=\mu_\text{ps}=\mu_\text{sp}=1.$

The results emphasize a number of points listed as follows:
\begin{itemize}
\item INR outperforms the RTD scheme in terms of both the throughput (Fig. 1a) and the outage probability (Fig. 1b).
\item Depending on the fading pdfs, HARQ does not necessarily increase the system throughput. For Rayleigh fading channels in particular, the HARQ-based throughput with the continuous (bursting) communication model is higher (lower) than the throughput when no HARQ is considered (Fig. 1a). The intuition behind the better system performance in the continuous model is that the good channel conditions are more efficiently exploited in this model. Particularly, using Jensen's inequality, convexity of $f(x)=\frac{1}{x}$, (7), (13) and (14) for, e.g., the INR protocol, we have
    \vspace{-2mm}
     \begin{align}
     \begin{array}{l}
 {\eta _{\text{continuous}}} \ge \frac{{D{{(1 - \Pr \{ \text{outage}\} )}^2}}}{{\sum_{m = 1}^{M + 1} {(\sum_{n = 1}^m {{l_n}} )\Pr \{ {A_m}\} } }} \\
  > \frac{{D{{(1 - \Pr \{ \text{outage}\} )}^2}}}{{\bar l}} = (1 - \Pr \{ \text{outage}\} ){\eta _{\text{bursting}}} \\
 \end{array}\nonumber
\end{align}
which emphasizes the validity of the argument as the outage probability vanishes, for instance when the number of (re)transmission rounds increases. (The same inequality can be written for the RTD protocol.) Finally, although not seen in the figures, the same conclusion is valid when, using the closed form expressions of the cdfs, e.g., (15), the transmission rates are optimized, in terms of throughput, via (4), (11), (13) and (14).
\item In all conditions, substantial outage probability reduction is achieved with limited number of retransmission rounds (Fig. 1b). Thus, the implementation of HARQ is more meaningful when the goal is to reduce the channel outage probability.
\item With imperfect SU-PU CSI, the PU tolerance, modeled by the probability constraint $\pi$, plays a great role in the SU-SU channel throughput; with relaxed PU received interference constraints (small $\pi$'s) the system throughput increases. However, the more secure the interference constraint should be satisfied, the less throughput is achieved at the secondary channel, converging to zero (Fig. 2a).
    \item The throughput difference between the bursting and continuous models diminishes under hard PU received interference power constraints, i.e., when $I_\text{p}$ decreases (Fig. 2a).
\item The throughput is more affected by the SU peak transmission power constraint as the PU transmission power increases, i.e., when the SU received SINR decreases (Fig.2b).
\end{itemize}
\begin{figure}
\vspace{-6mm}
\centering
  \includegraphics[width=0.83\columnwidth]{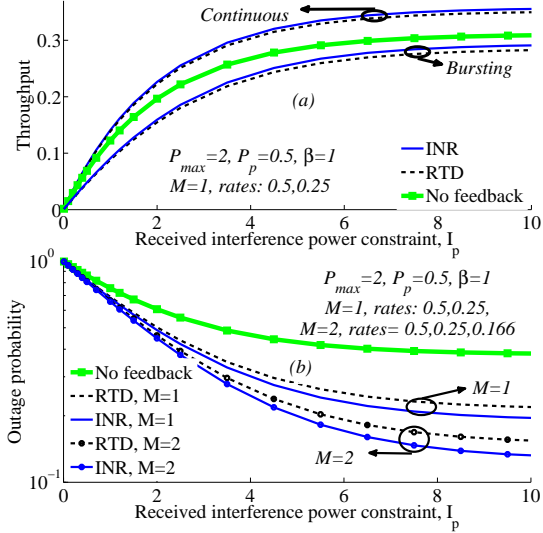}\\\vspace{-2mm}
\caption{(a) Throughput and (b) outage probability vs PU received interference power constraint $I_\text{p}$, perfect SU-PU CSI ($\beta=1$), $P_\text{max}=2,\, P_\text{p}=0.5$.}\label{figure111}
\vspace{-3mm}
\end{figure}
\vspace{-1mm}
\begin{figure}
\vspace{-0mm}
\centering
  \includegraphics[width=0.83\columnwidth]{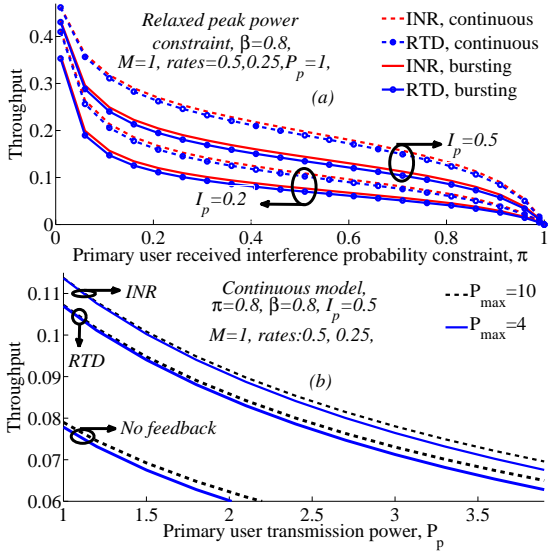}\\\vspace{-4mm}
\caption{Throughput vs (a) PU received interference probability constraint $\pi$ and (b) PU transmission power $P_\text{p}$, imperfect SU-PU CSI ($\beta=0.8$), a maximum of $M=1$ retransmission round. Figures 2a and 2b are obtained for relaxed SU peak power constraint and continuous communication model, respectively. }\label{figure111}
\vspace{-7mm}
\end{figure}
\vspace{-2mm}
\section{Conclusion}
\vspace{-0mm}
This letter studies the effect of HARQ on the performance of spectrum sharing networks. The results are obtained under bursting and continuous communication models when the SU is provided with imperfect SU-PU CSI. Under SU peak transmission power and PU received interference power constraints, the SU-SU channel throughput and the SU outage probability are determined for the INR and the RTD HARQ protocols. The results show that, although implementing HARQ protocols does not necessarily increase the system throughput in Rayleigh fading channels, considerable outage probability reduction is achieved in various conditions. Moreover, with imperfect interference information available at the SU transmitter, the PU tolerance significantly affects the SU performance. For different PU interference and SU peak transmission power constraints, the INR protocol outperforms the RTD scheme in terms of both the throughput and the outage probability. Also, higher rates are obtained in the continuous communication model when compared with the bursting communication model.
\vspace{-5mm}
\appendices
\section{Calculating the cdfs ${F_\Omega }(x) $ and ${F_{\phi_\text{p}} }(x|I_\text{p})$}
Define the auxiliary random variable $Z \buildrel\textstyle.\over= {P_\text{s}}{g_{\text{ss}}}$. Since ${P_\text{s}} = \min (P_\text{max},\frac{{{I_\text{p}}}}{{{{\tilde g}_{\text{sp}}}}})$, the cdf of $Z$ is found as
\vspace{-3mm}
\begin{align}
\begin{array}{l}
 {F_Z}(z) = 1 - \Pr \left\{ {g_{\text{ss}}} > \frac{z}{{{P_{\max }}}}\,\& \,{g_{\text{ss}}} > \frac{{z{{\tilde g}_{\text{sp}}}}}{{{I_\text{p}}}}\,\right\}  \\
  = 1 - \Pr \{ {{\tilde g}_{\text{sp}}} \le \frac{{{I_\text{p}}}}{{{P_{\max }}}}\} \Pr \{ {g_{\text{ss}}} > \frac{z}{{{P_{\max }}}}\,\}  \\
  - \int_{\frac{{{I_\text{p}}}}{{{P_{\max }}}}}^\infty  {{f_{{{\tilde g}_{\text{sp}}}}}(y)} \left(1 - {F_{{g_{\text{ss}}}}}(\frac{{zy}}{{{I_\text{p}}}})\right)\text{d}y \\
   \mathop  = \limits^{(a)} 1 - {e^{ - \frac{z}{{{\mu _{\text{ss}}}{P_{\max }}}}}}(1 - {e^{ - \frac{{{I_\text{p}}}}{{{\mu _{\text{sp}}}{P_{\max }}}}}}) - \frac{{{e^{ - (\frac{{{I_\text{p}}}}{{{\mu _{\text{sp}}}{P_{\max }}}} + \frac{z}{{{\mu _{\text{ss}}}{P_{\max }}}})}}}}{{1 + \frac{{{\mu _{\text{sp}}}}}{{{\mu _{\text{ss}}}{I_\text{p}}}}z}} \\
 \end{array}
\end{align}
where $(a)$ is based on the fact that for Rayleigh fading channels and the considered CSI imperfection model, (1), we have ${f_{{\tilde g_\text{sp}}}}(x) = \frac{1}{{{\mu _\text{sp}}}}{e^{ - \frac{x}{{{\mu _\text{sp}}}}}}$ \cite{cogasli,compressedneweurasipkhodemun,cog2}. In this way, from (17), ${F_\Omega }(x) = \int_0^\infty  {{f_{{g_{\text{ps}}}}}(z){F_Z}(x({P_\text{p}}z + {N_0})\text{d}z}$ and the definition of the incomplete Gamma function $\Gamma (x,y) = \int_y^\infty  {{u^{x - 1}}{e^{ - u}}\text{d}u}$, the cdf ${F_\Omega }(x)$ is found as stated in (15).

Further, the interference cdf ${F_{\phi_\text{p}} }(x|I_\text{p})$ can be written as
\vspace{-2mm}
\begin{align}
\begin{array}{l}
 {F_{{\phi _{\text{p}}}}}(x|{I_{\text{p}}}) = \Pr \{ {P_\text{s}}{g_{\text{sp}}} \le x\}  \\
  \,\,\,\,\,\,\,\,\,\,\,\,\,\,\,\,\,\,\,\,\,\,\,\,\,= 1 - \Pr \{ {g_{\text{sp}}} \ge \frac{x}{{{P_{\max }}}}\,\& \,{{\tilde g}_{\text{sp}}} \le \frac{{{I_\text{p}}{g_{\text{sp}}}}}{x}\}  \\
  \,\,\,\,\,\,\,\,\,\,\,\,\,\,\,\,\,\,\,\,\,\,\,\,\,= 1 - \int_{\frac{x}{{{P_{\max }}}}}^\infty  {\int_0^{\frac{{{I_\text{p}}u}}{x}} {{f_{{g_{\text{sp}}},{{\tilde g}_{\text{sp}}}}}(u,v)\text{d}u\text{d}v} }.  \\
 \end{array}
\end{align}
Here, ${{f_{{g_{\text{sp}}},{{\tilde g}_{\text{sp}}}}}}$ is the joint pdf of the variables ${{ g}_{\text{sp}}}$ and ${{\tilde g}_{\text{sp}}}$ which, using (1) and simple variable transformations, is found as
\vspace{-2mm}
\begin{align}
{f_{{g_{\text{sp}}},{{\tilde g}_{\text{sp}}}}}(y,z) = \frac{{{e^{ - \frac{{y + z}}{{(1 - {\beta ^2}){\mu _{\text{sp}}}}}}}}}{{(1 - {\beta ^2})\mu _{\text{sp}}^2}}{I_0}(\frac{{2\beta \sqrt {yz} }}{{(1 - {\beta ^2}){\mu _{\text{sp}}}}})
\end{align}
where $I_0$ is the zeroth-order modified Bessel function of the first kind \cite{cogasli,marcumqequations,846501,compressedneweurasipkhodemun,cog2}. Therefore, (18) is rephrased as
\vspace{-2mm}
\begin{align}
\begin{array}{l}
 {F_{{\phi _{\text{p}}}}}(x|{I_{\text{p}}}) = 1 - \frac{1}{{(1 - {\beta ^2})\mu _{\text{sp}}^2}}\int_{\frac{x}{{{P_{\max }}}}}^\infty  {{e^{ - \frac{y}{{(1 - {\beta ^2}){\mu _{\text{sp}}}}}}}\text{d}y}  \\
  \,\,\,\,\,\,\,\,\,\,\,\,\,\,\,\,\,\,\,\,\,\,\,\,\,\,\,\,\,\,\,\,\,\,\,\,\,\,\,\,\,\,\,\,\,\,\,\,\,\,\,\,\,\,\,\,\,\,\,\,\,\,\,\,\,\,\,\,\,\times \int_0^{\frac{{{I_\text{p}}y}}{x}} {{e^{ - \frac{z}{{(1 - {\beta ^2}){\mu _{\text{sp}}}}}}}{I_0}(\frac{{2\beta \sqrt {yz} }}{{(1 - {\beta ^2}){\mu _{\text{sp}}}}})\text{d}z}  \\
 \mathop  = \limits^{(b)} 1 - {e^{ - \frac{x}{{{\mu _{\text{sp}}}{P_{\max }}}}}} \\
  + \frac{1}{{{\mu _{\text{sp}}}}}\int_{\frac{x}{{{P_{\max }}}}}^\infty  {{e^{ - \frac{y}{{{\mu _{\text{sp}}}}}}}Q\left(\sqrt {\frac{{2y}}{{(1 - {\beta ^2}){\mu _{\text{sp}}}}}} \beta ,\sqrt {\frac{{2{I_\text{p}}y}}{{(1 - {\beta ^2}){\mu _{\text{sp}}}x}}} \right)\text{d}y}  \\
 \mathop  = \limits^{(c)} 1 - {e^{ - \frac{x}{{{\mu _{{\text{sp}}}}{P_{{\text{max}}}}}}}} + \frac{t}{r}Q\left(\sqrt {\frac{{(u - r)x}}{{2{P_{{\text{max}}}}}}} ,\sqrt {\frac{{(u + r)x}}{{2{P_{{\text{max}}}}}}} \right) \\
  + {e^{ - \frac{x}{{{\mu _{{\text{sp}}}}{P_{{\text{max}}}}}}}}Q\left(\beta \sqrt {\frac{{2x}}{{{P_{{\text{max}}}}w}}} ,\sqrt {\frac{{2{I_{\text{p}}}}}{{{P_{{\text{max}}}}w}}} \right) - \frac{1}{2}(1 + \frac{t}{r}){e^{\frac{{ - ux}}{{2{P_{{\text{max}}}}}}}}{I_0}(\frac{{2\beta \sqrt {x{I_{\text{p}}}} }}{{{P_{{\text{max}}}}w}}) \\
 w = {\mu _{{\text{sp}}}}(1 - {\beta ^2}),\,u = \frac{2}{{{\mu _{{\text{sp}}}}}}(1 + \frac{{{\beta ^2}{\mu _{{\text{sp}}}}}}{w} + \frac{{{I_{\text{p}}}{\mu _{{\text{sp}}}}}}{{xw}}) \\
 t = u - \frac{{4{I_{\text{p}}}}}{{wx}},\,r = \sqrt {{u^2} - \frac{{16{\beta ^2}{I_{\text{p}}}}}{{x{w^2}}}}.  \\
 \end{array}\nonumber
\end{align}
Again, $(b)$ is obtained by variable transform $\theta=\sqrt{z}$ and the definition of the Marcum Q-function $Q(a,b) = \int_b^\infty  {y{e^{\frac{{{y^2} + {a^2}}}{2}}}{I_0}(ay)\text{d}y}$. Finally, $(c)$ follows from variable transform $\xi=\sqrt{y}$ and \cite[eq. (55)]{marcumqequations}.
\vspace{-3mm}
\bibliographystyle{IEEEtran} 
\bibliography{masterICC2}
%
%
%
%

\vfill

\end{document}